\documentclass[prl,twocolumn,showpacs,amsmath,amssymb,superscriptaddress]{revtex4}
\usepackage{amssymb}
\usepackage{txfonts}
\usepackage{bbm}
\usepackage{graphicx}
\usepackage{appendix}
\usepackage{epsf}
\usepackage{epstopdf}
\usepackage{amsmath}
\usepackage[usenames]{color}

\begin{document}
\title{Impact of Step Defects on Surface States of Topological Insulators}


\author{Degang Zhang}
\affiliation{Texas Center for Superconductivity and Department of
Physics, University of Houston, Houston, TX 77204, USA}
\affiliation{Institute of Solid State Physics, Sichuan Normal
University, Chengdu 610066, China}
\author{C. S. Ting}
\affiliation{Texas Center for Superconductivity and Department of
Physics, University of Houston, Houston, TX 77204, USA}

\begin{abstract}

The eigenstates in the presence of a step defect (SD) along $x$ or
$y$ axis on the surface of topological insulators are exactly
solved. It is shown that unlike the electronic states in
conventional metals, the topological surface states across the SD
can produce spin rotations. The magnitudes of the spin rotations
depend on the height and direction of the SD. The oscillations of
local density of states (LDOS) are characterized by a wave vector
connecting two points on the hexagonal constant-energy contour at
higher energies. The period of the oscillation caused by the SD
along $y$ axis is $\sqrt 3$ ($\frac{1}{\sqrt 3}$) times that induced
by the SD along $x$ axis at a larger positive (negative) bias
voltage. With increasing the bias voltage, the period of the
oscillation, insensitive to the strength of the SD, becomes smaller.
At lower energies near the Fermi surface, the two types of wave
vectors coexist in the LDOS modulations. These results are
consistent qualitatively with recent observations of scanning
tunneling microscopy.

\end{abstract}


\maketitle

Recently topological surface states have attracted much attention in
the condensed matter community due to their potential applications
in quantum computing or spintronics [1,2]. The novel electronic
states, which preserve time-reversal symmetry, are produced by
spin-orbit interactions. Such Dirac-cone-like surface states have
been observed in three dimensional bulk insulating materials, such
as Bi$_{1-x}$Sb$_{x}$ [3], Bi$_2$Sb$_3$ [4], Sb$_2$Te$_3$ [5],
Bi$_2$Te$_3$ [5,6], TlBiSe$_2$ and TlBiTe$_2$ [7], by angle-resolved
photoemission spectroscopy (ARPES). The surface energy band
structure was determined by employing $k\cdot p$ theory [8], where
an unconventional hexagonal warping term plays a crucial role in
explaining the ARPES observations.

Scanning tunneling microscopy (STM) experiments have probed the
electronic waves in the presence of step defect (SD) on the surface
of the topological insulators Bi$_2$Te$_3$ [9,10] and the antimony
(Sb) [11,12]. The absence of backscattering of the topological
surface states makes the local density of states (LDOS) near the SD
more extraordinary as compared to that on the surface of
conventional metals [13,14]. In Ref. [10], Alpichshev et al.
observed the oscillations of the LDOS near a SD, dispersing with a
wave vector that may result from a hexagonal warping term. With
increasing the bias voltage, the period of the LDOS modulation
decreases. In this work, we investigate electron transport in the
presence of a SD along $x$ or $y$ axis on the surface of topological
insulators in the framework of quantum mechanics in order to explain
the STM experiments. We treat the SD as a $\delta(y)$ or $\delta(x)$
potential barrier, similar to that in conventional metals [13,14].
We note that in Ref. [15], the authors studied the scattering from a
$\delta(x)$ in strong topological insulators. However, they didn't
take the hexagonal warping term into account, which is the key to
understand the STM observations [9-12].

The momentum space Hamiltonian describing the surface states of
topological insulators reads [8]

$$H=(\frac{k^2}{2m^*}-\mu)I+v(k_x\sigma_y-k_y\sigma_x) +\lambda \phi(k_x,k_y)\sigma_z,
\eqno{(1)}
$$
where $I$ and $\sigma_i (i=x,y,z)$ are the $2\times 2$ unit matrix
and the Pauli matrices, respectively, $m^*$ is the effective mass of
electrons, which is usually very large for the topological
insulators, $\mu$ is the chemical potential, $v$ is the strength of
the Rashba spin-orbit coupling, the last term is the so called
hexagonal warping term, and $\phi(k_x,k_y)=k_x(k_x^2-3k_y^2)$. We
note that the real space Hamiltonian corresponding to Eq. (1)  can
be obtained by taking the transformations:
$k_x=-i\frac{\partial}{\partial x}$ and
$k_y=-i\frac{\partial}{\partial y}$. Obviously, the Hamiltonian (1)
has the eigenenergies
$$E_{{\bf k}s}=\frac{k^2}{2m^*}+(-1)^s{\cal E}_{\bf
k}-\mu \eqno{(2)}$$
with $s=0$ and $1$, and ${\cal E}_{\bf
k}=\sqrt{\lambda^2\phi^2(k_x,k_y)+v^2k^2}$.

We can see easily  that $x$ and $y$ directions in the Hamiltonian
(1) are inequivalent due to the existence of the warping term.
Therefore, it is naturally expected that different orientation of SD
leads to different oscillatory features in the LDOS. In the
following we study two special SDs, which are usually observed in
STM experiments.

{\it SD along $y$ axis}. We first consider the SD along $y$ axis,
which can be described by the $\delta$ potential $U(x)=U_0\delta(x)$
[13,14]. Here $U_0$ is the strength of the SD and is usually weak,
but $m^*U_0$ has a finite value.

\begin{figure}

\rotatebox[origin=c]{0}{\includegraphics[angle=0,
           height=1.4in]{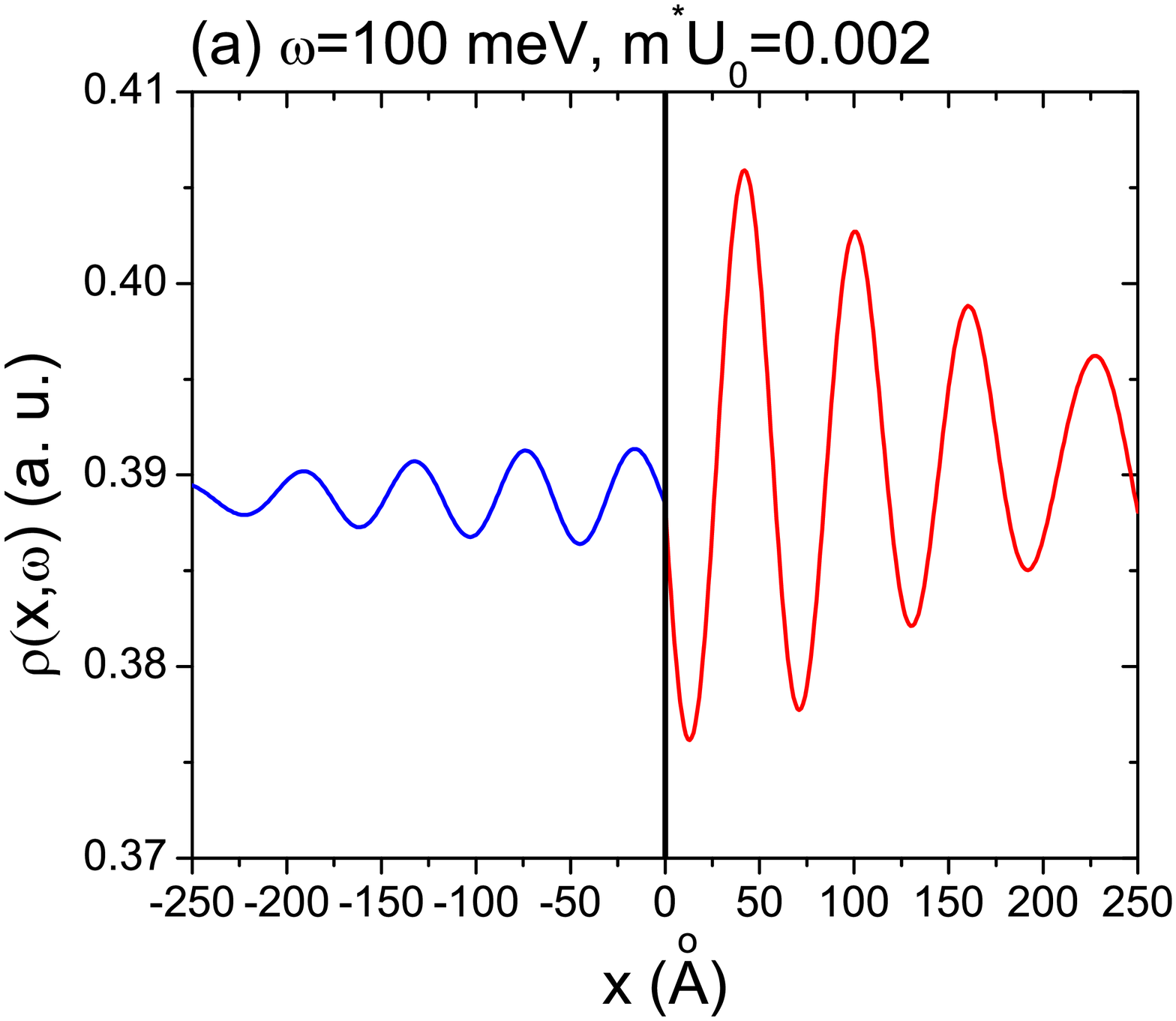}}
\rotatebox[origin=c]{0}{\includegraphics[angle=0,
           height=1.4in]{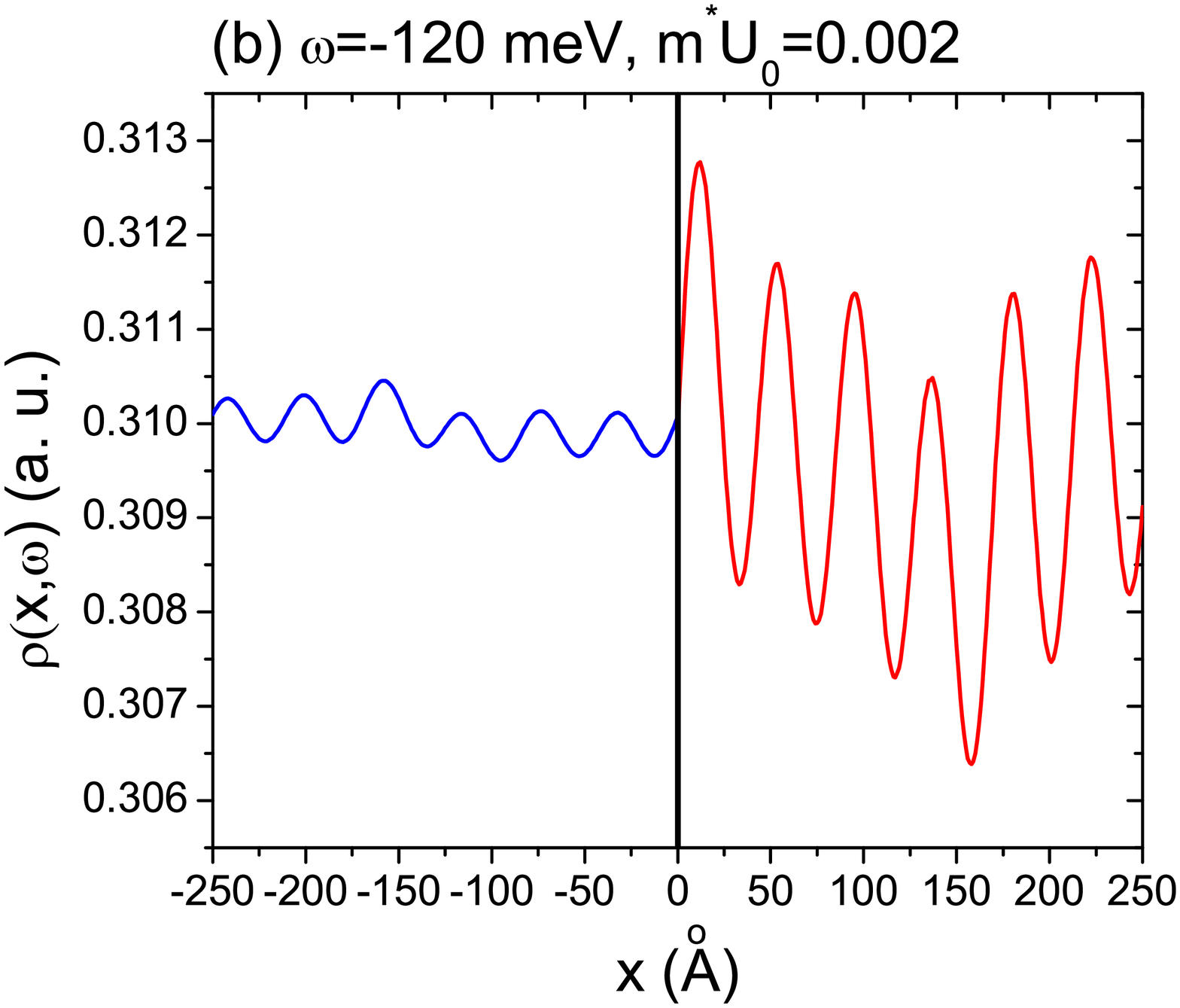}}
\rotatebox[origin=c]{0}{\includegraphics[angle=0,
           height=1.4in]{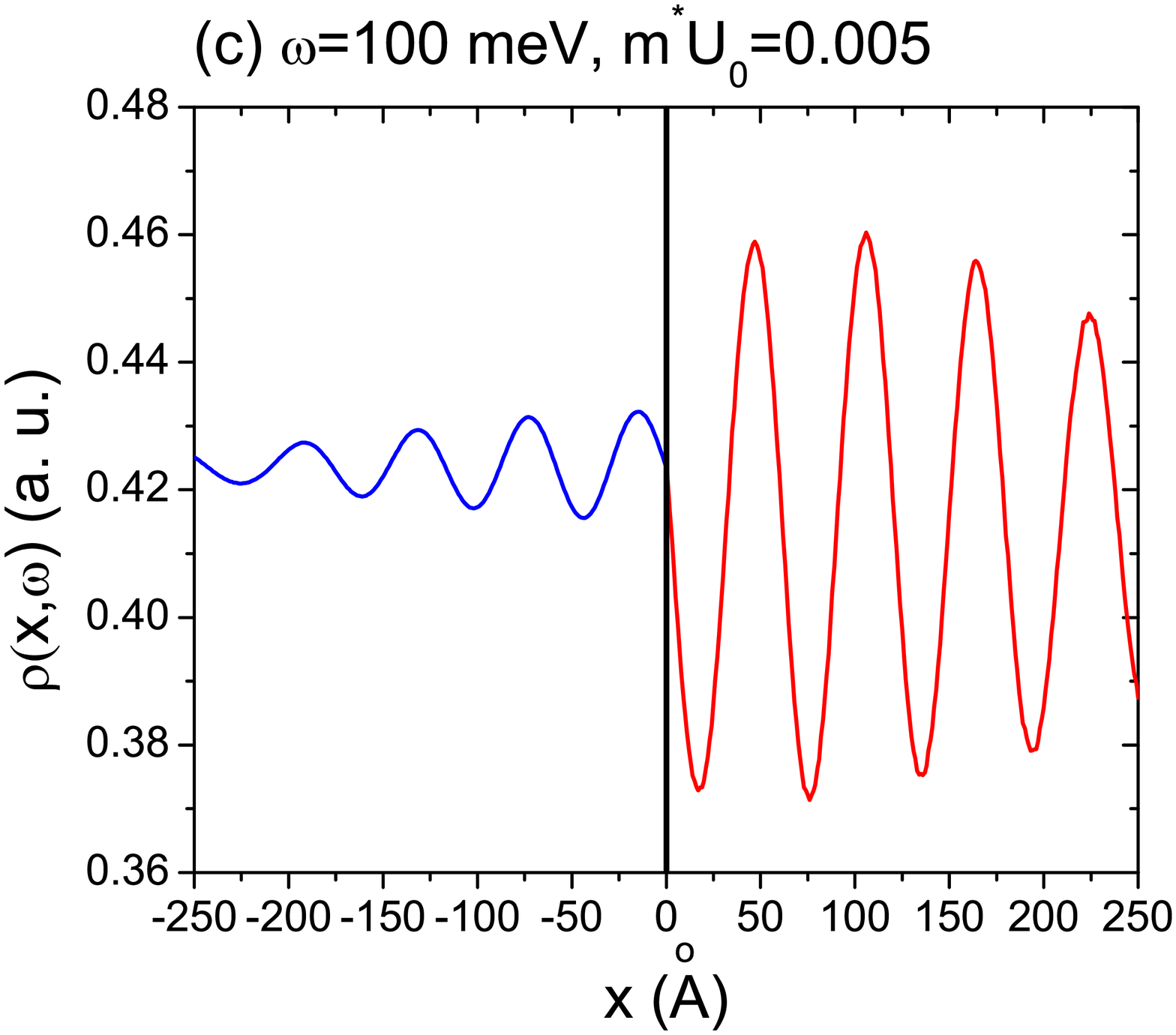}}
\rotatebox[origin=c]{0}{\includegraphics[angle=0,
           height=1.4in]{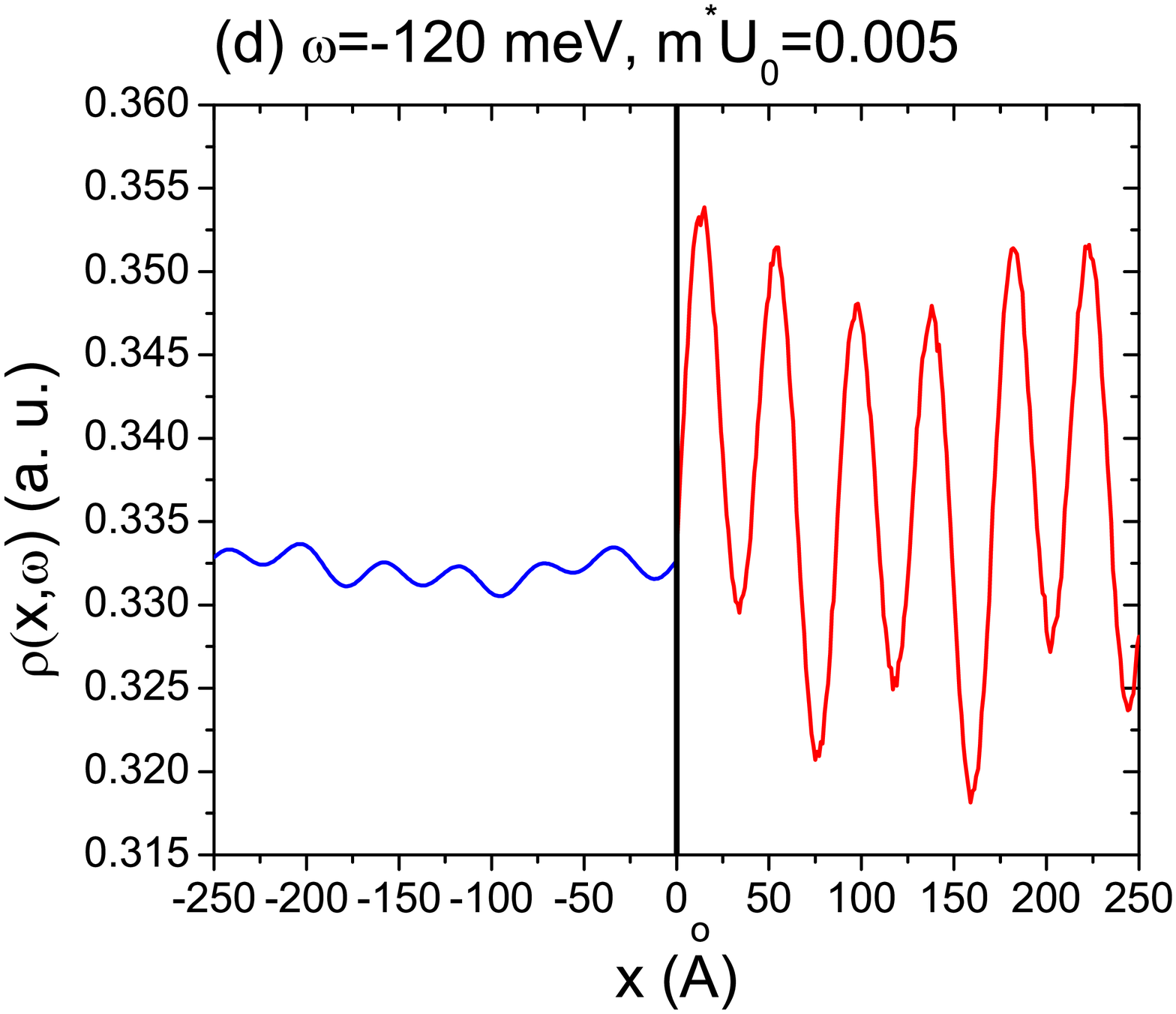}}
\caption{(Color online) The LDOS $\rho_{I,II}(x,\omega)$ as a
function of distance from the line defect along $y$ axis with
different values of $m^*U_0$ and bias voltages.}
\end{figure}

The wave function at the left of the SD, i.e. $x<0$, is

$$\psi_{I}^s(x,y;{\bf k})=\frac{e^{i(k_xx+k_yy)}}{\Gamma_s(k_x,k_y)}\left(
\begin{array}{c}
\xi_s(k_x,k_y)\\v(ik_x-k_y) \end{array} \right) $$
$$+R_s\cdot \frac{e^{i(-k_xx+k_yy)}}{\Gamma_s(-k_x,k_y)}\left(
\begin{array}{c}
\xi_s(-k_x,k_y)\\v(-ik_x-k_y) \end{array} \right), \eqno{(3)}
$$
where $\xi_s(k_x,k_y)=\lambda \phi(k_x,k_y)+(-1)^s{\cal E}_{\bf k},$
$\Gamma_s(k_x,k_y)=\sqrt{\xi_s^2(k_x,k_y)+v^2k^2}$, the first and
second terms represent the incoming and reflection wave functions,
respectively, while the outcoming wave function at the right of the
SD, i.e. $x>0$, is found to have general form

$$\psi_{II}^s(x,y;{\bf k})=T_s\cdot \frac{e^{i(k_xx+k_yy)}}{\Gamma_s(k_x,k_y)}\left(
\begin{array}{c}
\xi_s(k_x,k_y)\\v(ik_x-k_y) \end{array} \right) $$
$$+e^{ik_yy}\left(
\begin{array}{c}
A_s\cos(k_xx)+B_s\sin(k_xx)\\C_s\cos(k_xx)+D_s\sin(k_xx) \end{array}
\right), \eqno{(4)}
$$
where $C_s$ and $D_s$ are arbitrary constants to be determined by
the boundary conditions at the SD, i.e. $x=0$, and

$$A_s=-u_sC_s-v_sD_s, ~~~~~B_s=v_sC_s-u_sD_s,$$
$$u_s=(vk^2)^{-1}[i\lambda k_x\phi(k_x,k_y)+(-1)^sk_y{\cal E}_{\bf
k}]$$
$$v_s=(vk^2)^{-1}[-i\lambda k_y\phi(k_x,k_y)+(-1)^sk_x{\cal E}_{\bf
k}].\eqno{(5)}$$

We note that the first term in Eq. (4) is the tunneling wave
function while the second term describes an extra spin rotation,
distinguishing from the electronic wave function in conventional
metals. It is nothing but this term that leads to the oscillations
of the LDOS at $x>0$ [11].

Integrating the coupled Schrodinger equations associated with the
Hamiltonian (1) in real space plus the $\delta(x)$ potential, we
have the boundary conditions at the SD,

$$\psi_{I}^s(0,y;{\bf k})=\psi_{II}^s(0,y;{\bf k})$$

$$\frac{\partial \psi_{II}^s(0,y;{\bf k})}{\partial x}-\frac{\partial \psi_{I}^s(0,y;{\bf k})}{\partial x}
=2m^*U_0\psi_{I}^s(0,y;{\bf k}).\eqno{(6)}$$ Substituting Eqs. (3)
and (4) into Eq. (6), we obtain the spin rotation constants

\begin{widetext}

$$
C_s=-\frac{m^*U_0\eta_s^+}{\Gamma_s(k_x,k_y)k_x(ik_x-m^*U_0)}\cdot
\frac{k_x[\xi_s(k_x,k_y)+u_sv(ik_x-k_y)]+v_s(ik_x-2m^*U_0)v(ik_x-k_y)}{v^2k^2(1+u_s^2+v_s^2)+u_s\eta_s^-+iv_s\eta_s^+},
$$

$$
D_s=-\frac{m^*U_0\eta_s^+}{\Gamma_s(k_x,k_y)k_x(ik_x-m^*U_0)}\cdot
\frac{v_sk_xv(ik_x-k_y)-(ik_x-2m^*U_0)[\xi_s(k_x,k_y)+u_sv(ik_x-k_y)]}{v^2k^2(1+u_s^2+v_s^2)+u_s\eta_s^-+iv_s\eta_s^+},\eqno{(7)}
$$

\end{widetext}
where $\eta_s^\pm=\xi_s(-k_x,k_y)v(ik_x-k_y)\pm
\xi_s(k_x,k_y)v(ik_x+k_y)$, and the reflection and tunneling
coefficients

$$R_s=\frac{\Gamma_s(-k_x,k_y)}{\eta_s^+}[A_sv(ik_x-k_y)-C_s\xi_s(k_x,k_y)],$$

$$T_s=1-\frac{\Gamma_s(k_x,k_y)}{\eta_s^+}[A_sv(ik_x+k_y)+C_s\xi_s(-k_x,k_y)].\eqno{(8)}$$
Obviously, when $U_0=0$, we have $C_s=D_s=R_s=0$ and $T_s=1$.

In order to compare with the STM experiments, we calculate the LDOS
near the SD, which can be expressed as

$$\rho_{I,II}(x,\omega)=\sum_{k_x>0,k_y,s}|\psi_{I,II}^s(x,y;{\bf k})|^2\delta(\omega-E_{{\bf
k}s}).\eqno{(9)}$$ Here we emphasize that the formula (9) only
considers the contributions of the topological surface states with
$k_x>0$ so that we can know clearly the LDOS modulations induced by
the incoming and reflection wave functions or the outcoming wave
function. We note that the LDOS observed by STM experiments should
also include the contributions coming from those surface states with
$k_x<0$, i.e. $\rho_{\rm
STM}(|x|,\omega)=\rho_{I}(x,\omega)+\rho_{II}(x,\omega)$. In our
following calculations, we use the physical parameters of the
topological insulator Bi$_2$Te$_3$: $\lambda=250.0$
eV$\cdot$\AA$^3$, $v=2.55$ eV$\cdot$\AA$^3$, and $\mu=0.334$ eV
[8,10]. In Fig. 1, we present the LDOS with different values of
$m^*U_0$ at high positive and negative energies. Obviously, the
amplitude of the LDOS modulation near the SD depends strongly on the
strength $U_0$ of the $\delta$ potential and the bias voltage
$\omega$. However, if $\omega$ is fixed, both period and phase of
the oscillation keep unchanged with increasing $U_0$. When
$\omega=100$ meV, the period $T_y(\omega)\approx 58.0$ \AA  in both
sides of the SD. When $\omega=-120$ meV, the period
$T_y(\omega)\approx 42.0$ \AA. At the positions with the same
distance from the SD, the LDOS at these bias voltages has a maximum
value and a minimum value, respectively.

In order to understand the oscillatory features of the LDOS, we plot
the constant-energy contours of the topological surface state band
at $\omega=100$ meV and $-120$ meV in Fig. 2. We observe that the
periods of the LDOS modulations at these energies are associated
with a wave vector connecting two points on the corresponding
constant-energy contours. In other words,
$T_y(100)\sim\frac{2\pi}{|{\bf q}_A-{\bf q}_B|}=\frac{2\pi}{|{\bf
q}_C-{\bf q}_D|}=\frac{\pi}{0.0549}=57.2$ \AA, and
$T_y(-120)\sim\frac{2\pi}{|{\bf q}_E-{\bf q}_F|}=\frac{2\pi}{|{\bf
q}_G-{\bf q}_H|}=\frac{\pi}{0.0727}=43.2$ \AA. Therefore, such
oscillations of the LDOS are due to quasiparticle interference. It
is obvious that there is no backscattering of the topological
surface states in the LDOS, which is associated with the wave vector
connecting two crossing points between the constant-energy contour
and the $x$ axis.

\begin{figure}

\rotatebox[origin=c]{0}{\includegraphics[angle=0,
           height=1.7in]{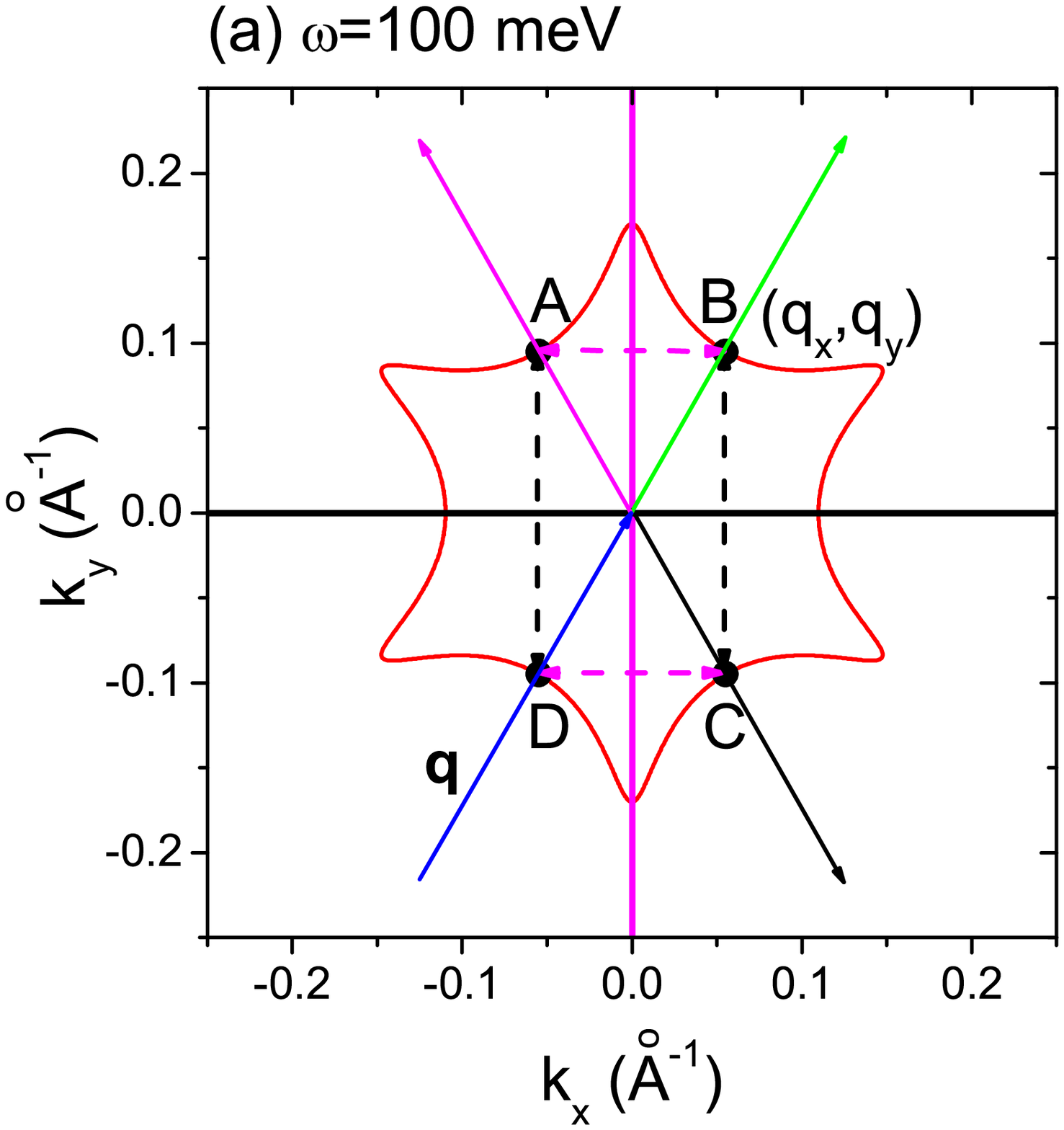}}
\rotatebox[origin=c]{0}{\includegraphics[angle=0,
           height=1.7in]{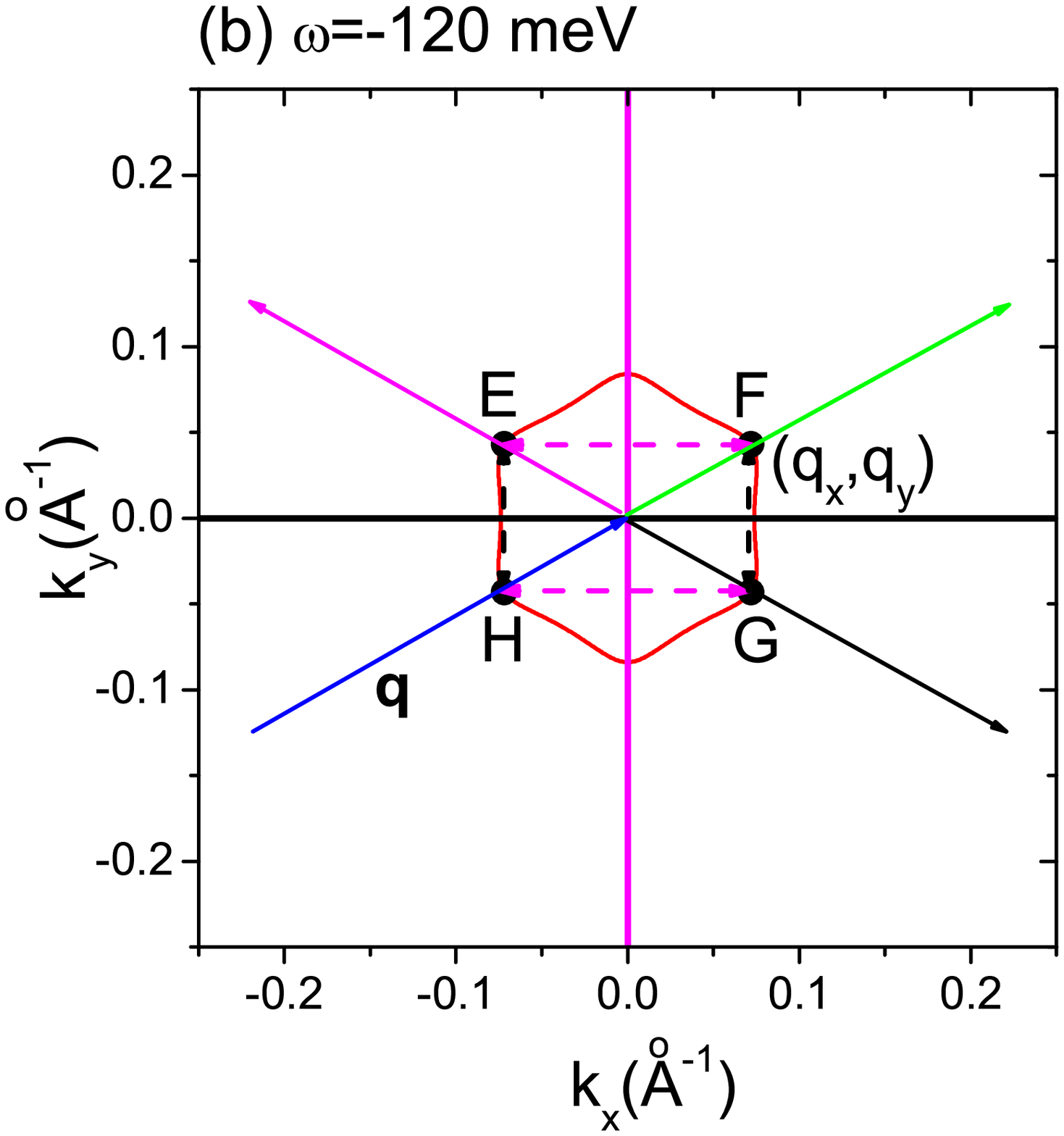}}
\caption{(Color online) The constant-energy contours of the surface
state band at different energies.}
\end{figure}

{\it SD along $x$ axis}. Now we consider the SD along $x$ axis,
corresponding to that observed in Bi$_2$Te$_3$ by the STM experiment
[10]. Similarly, the wave function at $y<0$ has the form

$$\psi_{I}^s(x,y;{\bf k})=\frac{e^{i(k_xx+k_yy)}}{\Gamma_s(k_x,k_y)}\left(
\begin{array}{c}
\xi_s(k_x,k_y)\\v(ik_x-k_y) \end{array} \right) $$
$$+{\cal R}_s\cdot \frac{e^{i(k_xx-k_yy)}}{\Gamma_s(k_x,k_y)}\left(
\begin{array}{c}
\xi_s(k_x,k_y)\\v(ik_x+k_y) \end{array} \right), \eqno{(10)}
$$
where we have used $\xi_s(k_x,-k_y)\equiv\xi_s(k_x,k_y)$ and
$\Gamma_s(k_x,-k_y)\equiv\Gamma_s(k_x,k_y)$. The wave function at
$y>0$ is

$$\psi_{II}^s(x,y;{\bf k})={\cal T}_s\cdot \frac{e^{i(k_xx+k_yy)}}{\Gamma_s(k_x,k_y)}\left(
\begin{array}{c}
\xi_s(k_x,k_y)\\v(ik_x-k_y) \end{array} \right) $$
$$+e^{ik_xx}\left(
\begin{array}{c}
{\cal A}_s\cos(k_yy)+{\cal B}_s\sin(k_yy)\\{\cal
C}_s\cos(k_yy)+{\cal D}_s\sin(k_yy)
\end{array} \right), \eqno{(11)}
$$
where ${\cal A}_s=w_s(k_x{\cal C}_s-k_y{\cal D}_s), {\cal
B}_s=w_s(k_y{\cal C}_s+k_x{\cal D}_s),$
$w_s=(ivk^2)^{-1}\xi_s(k_x,k_y)$. The spin rotation constants ${\cal
C}_s$ and ${\cal D}_s$, the reflection coefficient ${\cal R}_s$, and
the tunneling coefficient ${\cal T}_s$ are determined by the
following constraints

$$\psi_{I}^s(x,0;{\bf k})=\psi_{II}^s(x,0;{\bf k})$$

$$(\frac{1}{2m^*}I-3\lambda k_x\sigma_z
)\left( \frac{\partial \psi_{II}^s(x,0;{\bf k})}{\partial
y}-\frac{\partial \psi_{I}^s(x,0;{\bf k})}{\partial y}\right)$$
$$=U_0\psi_{I}^s(x,0;{\bf k}).\eqno{(12)}$$
Solving Eq. (12), we have

$${\cal
C}_s=\frac{\zeta_s^-}{\Gamma_s(k_x,k_y)}\cdot
\frac{U^-\xi_s(k_x,k_y)a_s^{22}-U^+v(ik_x-k_y)a_s^{12}}{a_s^{11}a_s^{22}+a_s^{12}a_s^{21}},$$

$${\cal
D}_s=\frac{\zeta_s^-}{\Gamma_s(k_x,k_y)}\cdot
\frac{U^+v(ik_x-k_y)a_s^{11}+U^-\xi_s(k_x,k_y)a_s^{21}}{a_s^{11}a_s^{22}+a_s^{12}a_s^{21}},$$

$${\cal R}_s=\frac{\Gamma_s(k_x,k_y)}{\zeta_s^-}[{\cal A}_sv(ik_x-k_y)-{\cal C}_s\xi_s(k_x,k_y)],$$

$${\cal T}_s=1+\frac{\Gamma_s(k_x,k_y)}{\zeta_s^-}[{\cal A}_sv(ik_x+k_y)-{\cal C}_s\xi_s(k_x,k_y)],\eqno{(13)}$$
where $U^\pm=\frac{U_0}{\frac{1}{2m^*}\pm 3\lambda k_x}$,
$\zeta_s^\pm=\xi_s(k_x,k_y)[v(ik_x-k_y)\pm v(ik_x+k_y)]$, and

$$a_s^{11}=w_s\zeta_s^-(k_y^2-\frac{1}{2}k_xU^-)+(ik_y-\frac{1}{2}U^-)[w_sk_x\zeta_s^+-2\xi_s^2(k_x,k_y)],$$

$$a_s^{12}=w_sk_y[\zeta_s^-(k_x+\frac{1}{2}U^-)-\zeta_s^+(ik_y-\frac{1}{2}U^-)],$$

$$a_s^{21}=\frac{1}{2}U^+\zeta_s^-+(ik_y-\frac{1}{2}U^+)(\zeta_s^++2w_sk_xv^2k^2),$$

$$a_s^{22}=k_y[\zeta_s^-+w_sv^2k^2(2ik_y-U^+)].\eqno{(14)}$$

\begin{figure}

\rotatebox[origin=c]{0}{\includegraphics[angle=0,
           height=1.4in]{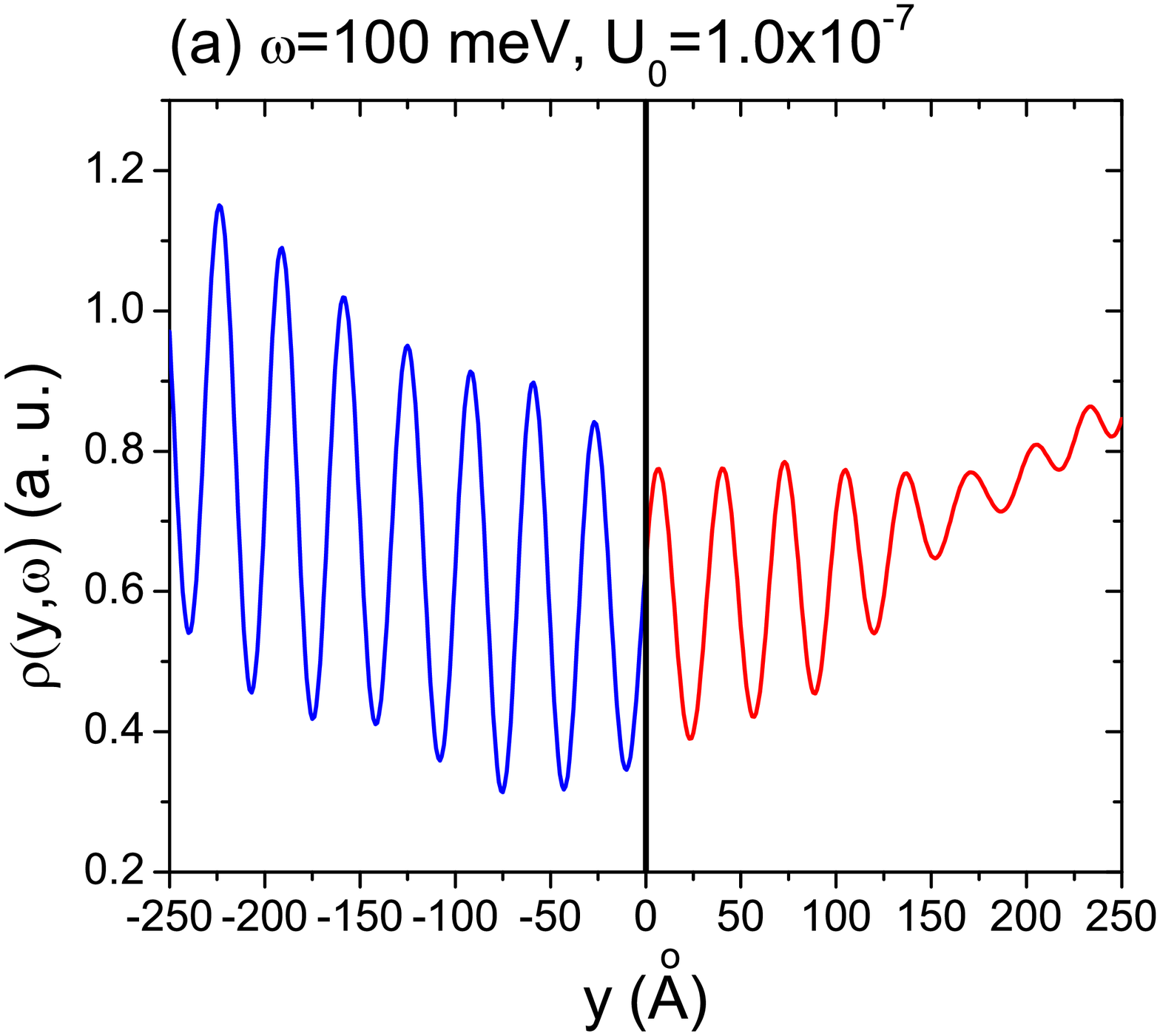}}
\rotatebox[origin=c]{0}{\includegraphics[angle=0,
           height=1.4in]{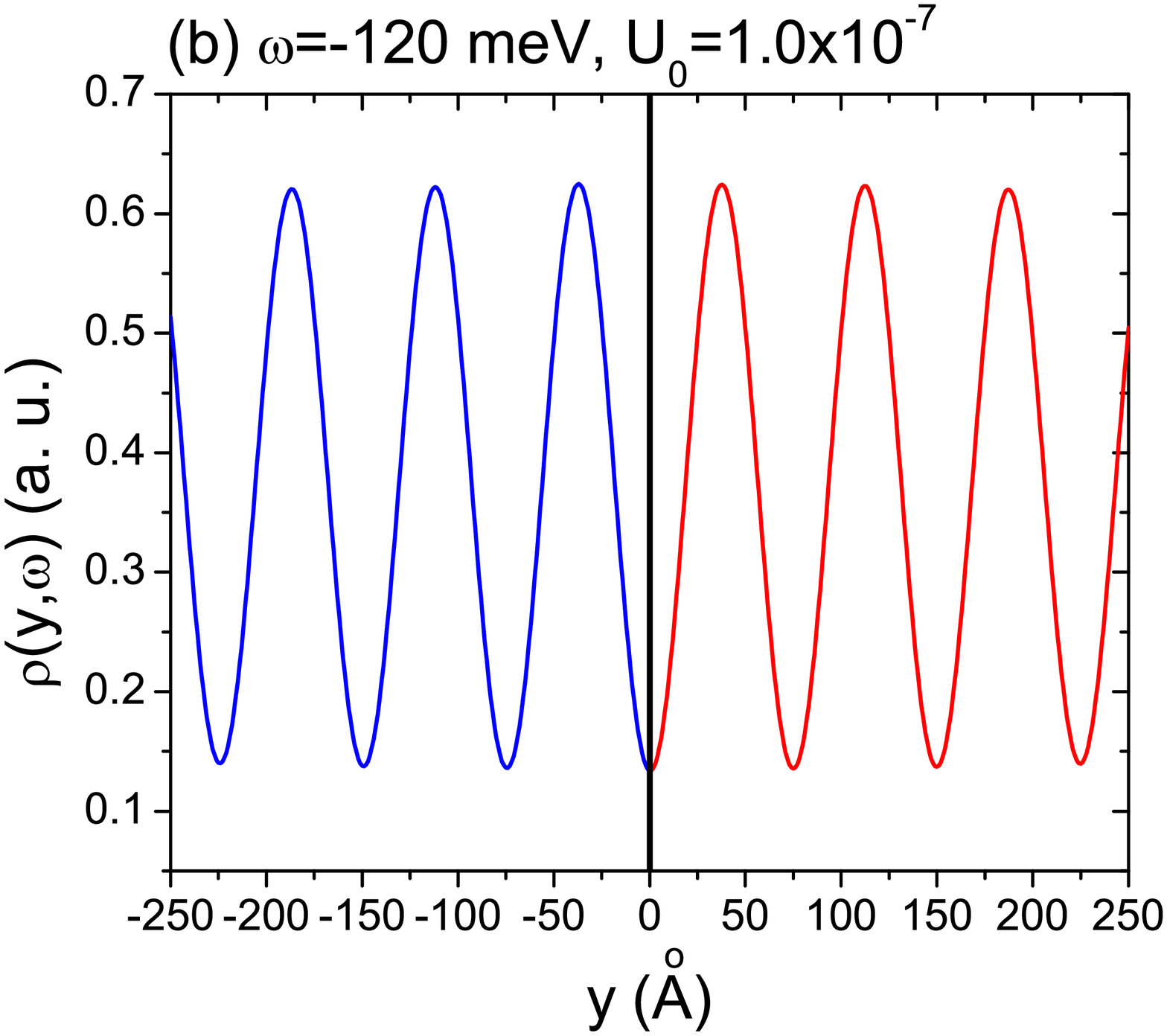}}
\rotatebox[origin=c]{0}{\includegraphics[angle=0,
           height=1.4in]{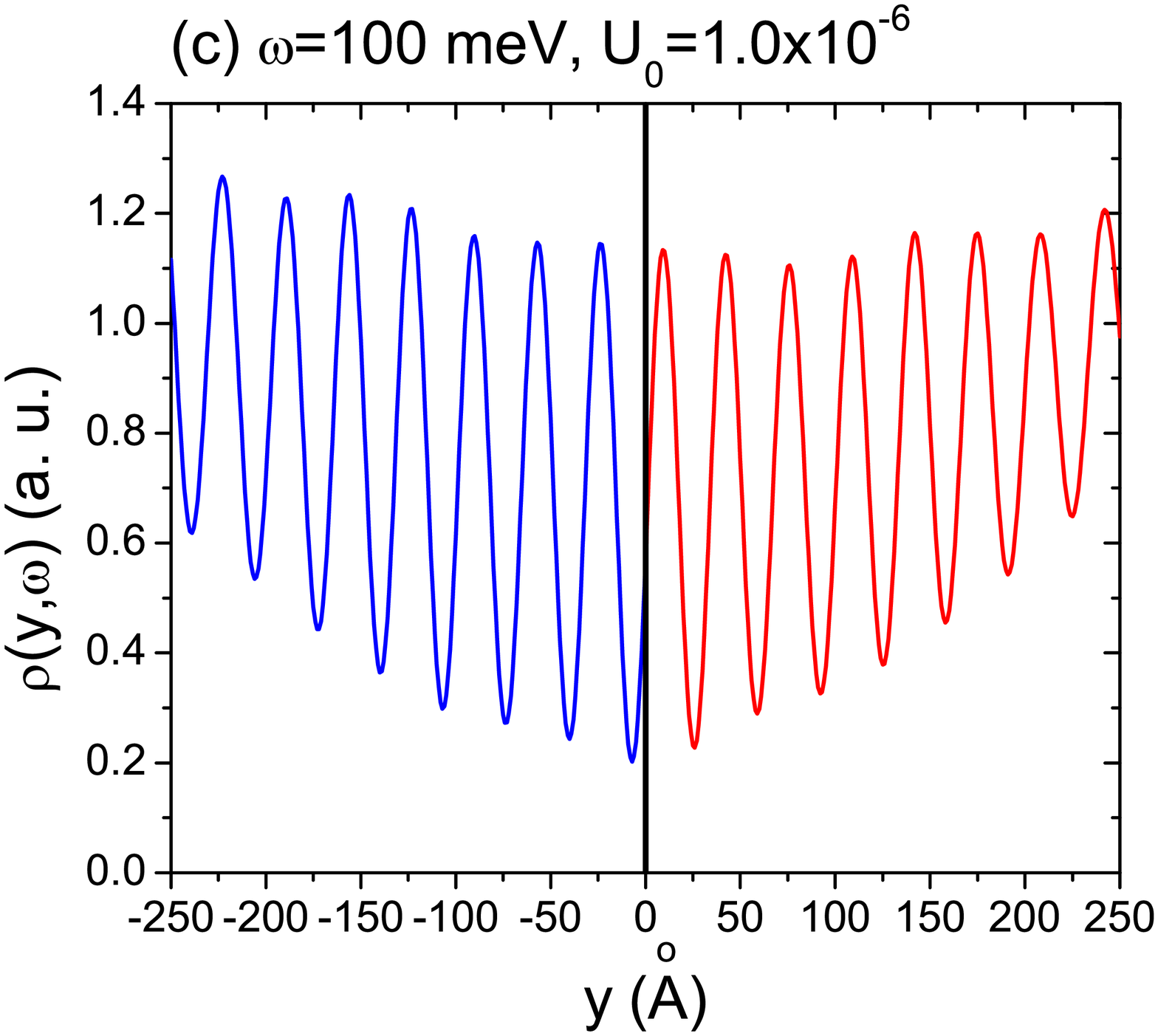}}
\rotatebox[origin=c]{0}{\includegraphics[angle=0,
           height=1.4in]{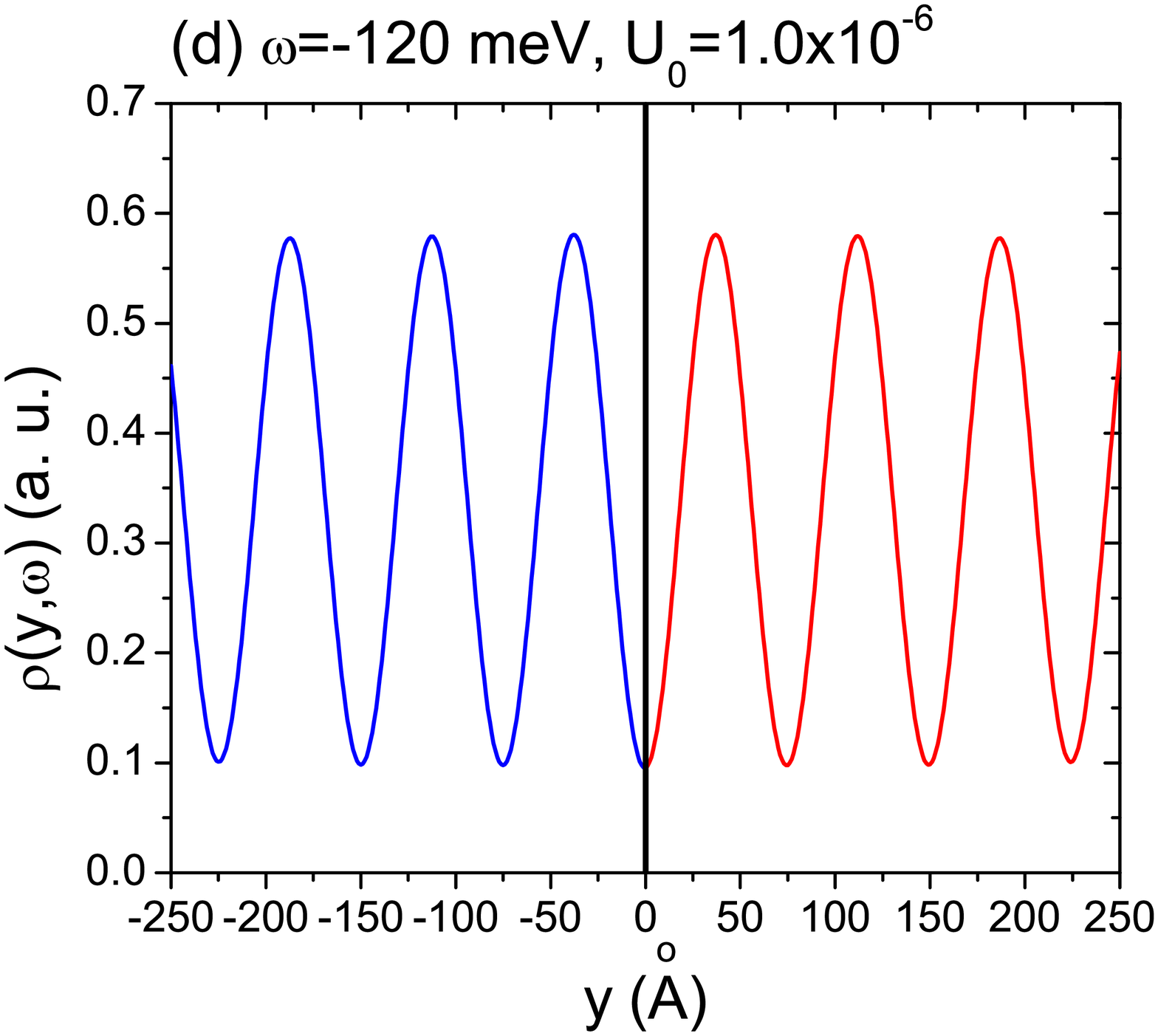}}
\caption{(Color online) The LDOS $\rho_{I,II}(y,\omega)$ as a
function of distance from the line defect along $x$ axis with
different values of $U_0$ and bias voltages.}
\end{figure}

\begin{figure}

\rotatebox[origin=c]{0}{\includegraphics[angle=0,
           height=1.4in]{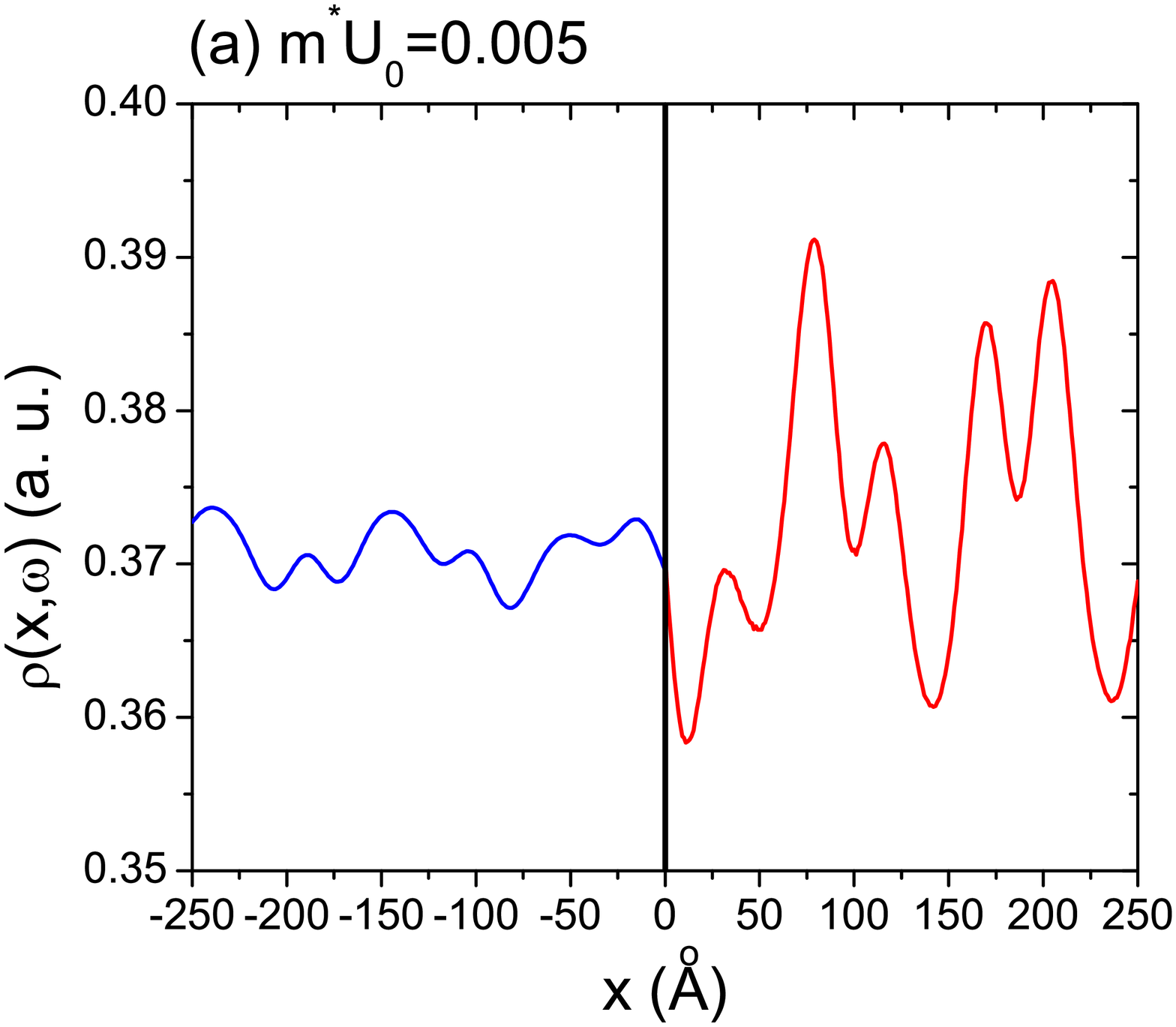}}
\rotatebox[origin=c]{0}{\includegraphics[angle=0,
           height=1.4in]{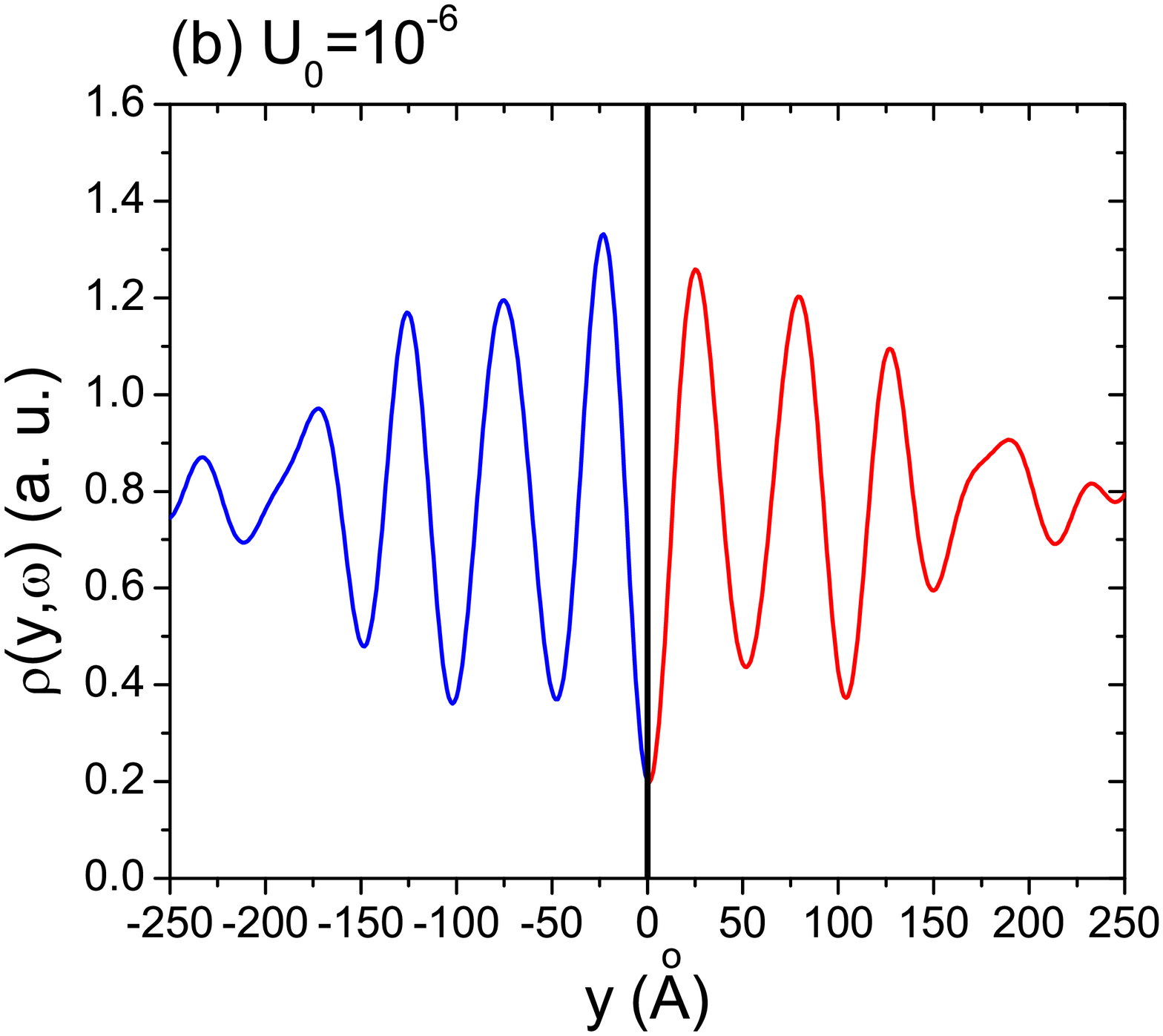}}
\caption{(Color online) The LDOS as a function of distance from the
line defect along $y$ or $x$ axis at zero bias voltage.}
\end{figure}

Correspondingly, the LDOS in this case is

$$\rho_{I,II}(y,\omega)=\sum_{k_x,k_y>0,s}|\psi_{I,II}^s(x,y;{\bf k})|^2\delta(\omega-E_{{\bf
k}s}).\eqno{(15)}$$ According to Eq. (15) and using the parameters
in Bi$_2$Te$_3$, we also calculate the LDOS near the SD along $x$
axis at different bias voltages and strengths of the $\delta$
potential, shown in Fig. 3. We can see that the amplitude of the
LDOS modulation also changes with $U_0$ and $\omega$. When $\omega$
is fixed, the period and the phase of the oscillations are also
independent of $U_0$. However, on the points symmetrical about the
SD, the LDOS at $\omega=100$ meV has the same oscillatory features
with that induced by the SD along $y$ axis. In contrast, when
$\omega=-120$ meV, the LDOS has two maximum or minimum values. We
note that $T_x(100)\approx 33.0$ \AA and $T_x(-120)\approx 74.0$
\AA. The oscillatory characteristics are also produced by
quasiparticle interference between two points on the constant-energy
contours in Fig. 2, similar to the previous case. We find that
$T_x(100)\sim \frac{2\pi}{|{\bf q}_B-{\bf q}_C|}=\frac{2\pi}{|{\bf
q}_A-{\bf q}_D|}=\frac{\pi}{0.0952}=33.0$ \AA while $T_x(-120)\sim
\frac{2\pi}{|{\bf q}_F-{\bf q}_G|}=\frac{2\pi}{|{\bf q}_E-{\bf
q}_H|}=\frac{\pi}{0.0419}=74.98$ \AA. With increasing $\omega$,
$|{\bf q}_B-{\bf q}_C|$ and $|{\bf q}_F-{\bf q}_G|$ become longer,
and so the periods of the oscillations become smaller. When $\omega$
decreases to the values near the Dirac point, the quasiparticle
interference associated with the wave vector ${\bf q}_E-{\bf q}_H$
or ${\bf q}_F-{\bf q}_G$ becomes very weaker and the periods of the
oscillations become very larger. Therefore, the LDOS is almost
constant. These results are consistent with the STM observations
[10,11].

Because the modulation wave vector at $\omega=100$ meV is different
from that at $\omega=-120$ meV, it is expected that the two
modulation wave vectors compete at small energies. Fig. 4 shows the
LDOS at zero bias voltage for the two kinds of SDs. Obviously, the
LDOS modulations cannot be fitted by a wave vector connecting two
points on the Fermi surface [10].

In summary, we have investigated the impact of the SD along $x$ or
$y$ axis on the surface states of topological insulators. We
discover for the first time  that there are spin rotations when the
topological surface states move through the $\delta$ potential
barrier. The oscillations of the LDOS near the SDs are induced by
quasiparticle interference. This agrees qualitatively with the STM
experiments. The period and phase of the oscillations are
independent of the strength of SDs at high positive or negative bias
voltage. But the amplitudes of the oscillations are sensitive to the
strength of SDs and the bias voltages. We also find that the
oscillations of the LDOS at high energies induced by the SD along
$y$ or $x$ axis are associated with the same points on the
constant-energy contours. Therefore, their periods have special
relations, i.e. $T_y(|\omega|)=\sqrt{3}T_x(|\omega|)$ and
$T_y(-|\omega|)=\frac{1}{\sqrt{3}}T_x(-|\omega|)$, where $\omega$ is
large. We hope that such relations could be verified by future STM
experiments.

This work was supported by the Texas Center for Superconductivity at
the University of Houston and by the Robert A. Welch Foundation
under the Grant no. E-1411.


\begin{thebibliography}{99}
\bibitem{1} Xiao-Liang Qi and Shou-Cheng Zhang, Phys. Today, {\bf 63}, 33 (2010); Rev. Mod. Phys. (in press).
\bibitem{2} M. Z. Hasan and C. L. Kane, Rev. Mod. Phys. {\bf 82}, 3045 (2010).
\bibitem{3} D. Hsieh {\it et al}, Science {\bf 323}, 919 (2009).
\bibitem{4} Y. Xia {\it et al}, Nature Phys. {\bf 5}, 398 (2009).
\bibitem{5} H. Zhang {\it et al}, Nature Phys. {\bf 5}, 438 (2009).
\bibitem{6} Y.L. Chen {\it et al}, Science {\bf 325}, 178 (2009).
\bibitem{7} Y.L. Chen {\it et al}, Phys. Rev. Lett. {\bf 104}, 016401 (2010).
\bibitem{8} Liang Fu, Phys. Rev. Lett. {\bf 105}, 266401 (2010).
\bibitem{9} T. Zhang {\it et al}, Phys. Rev. Lett. {\bf 103}, 266803 (2009).
\bibitem{10} Z. Alpichshev {\it et al}, Phys. Rev. Lett. {\bf 104}, 016401 (2010).
\bibitem{11} K. K. Gomes {\it et al}, arXiv:0909.0921 (unpublished).
\bibitem{12} J. Seo {\it et al}, Nature (London) {\bf 466}, 343 (2010).
\bibitem{13} M.F. Crommie, C. P. Lutz, and D. M. Eigler, Nature  {\bf 363}, 524 (1993).
\bibitem{14} L. C. Davis {\it et al}, Phys. Rev. B {\bf 43}, 3821 (1991).
\bibitem{15} R. R. Biswas and A. V. Balatsky, Phys. Rev. B {\bf 83}, 075439 (2011).


\end{thebibliography}
\end{document}